\documentclass[conference]{IEEEtran}
\IEEEoverridecommandlockouts

\usepackage{cite}
\usepackage{amsmath,amssymb,amsfonts}
\usepackage{algorithmic}
\usepackage{graphicx}
\usepackage{textcomp}

\begin{document}

\title{Lightweight Underwater Acoustic Vessel Classification for Embedded Deployment}

\author{
\IEEEauthorblockN{
Abishek Soti, Thura Pyae Sone, Naqib Ibnul, Jim Huang,\\
Sergio Chevtchenko, Henry Zhong, Gregory Cohen, Ying Xu
}
\IEEEauthorblockA{
\textit{ICNS, Western Sydney University}\\
Sydney, Australia\\[3pt]
\texttt{A.soti@westernsydney.edu.au},
\texttt{T.Sone@westernsydney.edu.au}\\
\texttt{n.ibnul@westernsydney.edu.au},
\texttt{J.Huang5@westernsydney.edu.au}\\
\texttt{S.Chevtchenko@westernsydney.edu.au},
\texttt{H.Zhong@westernsydney.edu.au}\\
\texttt{G.Cohen@westernsydney.edu.au},
\texttt{ying.xu@westernsydney.edu.au}
}
}

\maketitle

\begin{abstract}
We propose a Raspberry Pi-based underwater vessel classification system for edge deployment under limited connectivity, bandwidth and power in offshore acoustic surveillance. Using ShipsEar, we compare six classifiers on a Raspberry Pi 5 through a complete audio-to-prediction pipeline. We assess classification performance, processing latency, CPU utilisation, memory consumption and monitored power. Supporting experiments examine acoustic feature selection, statistical pooling and Transformer patch configurations. The two-layer convolutional neural network (CNN) achieved a mean macro-F1 of \textbf{(99.14 \textpm{} 0.32)\%} across five random seeds. The radial basis function support vector machine (RBF-SVM) achieved \textbf{98.83\%}, with a median end-to-end latency of 28.109~ms compared with 120.655~ms for the CNN. This favourable trade-off motivated RBF-SVM deployment on the Raspberry Pi Zero 2 W. It maintained the same classification performance with a median end-to-end latency of \textbf{349.09~ms} and a sustained throughput of \textbf{2.962 classifications/s}, or \textbf{14.8} times the rate required for consecutive five-second audio segments.
\end{abstract}

\vspace{0.5\baselineskip}


\section{Introduction}

Shore-connected passive acoustic monitoring systems enable continuous observation of underwater soundscapes, but their dependence on coastal infrastructure can restrict monitoring locations~\cite{pierpoint2026soundscapes}. Deploying hydrophones offshore can extend monitoring coverage, although transmitting acoustic recordings may be impractical due to limited connectivity, bandwidth and power~\cite{underwater_networks_pal2022}. Processing these recordings directly on embedded devices offers an alternative by enabling local vessel classification and reducing reliance on continuous transmission.

However, high classification accuracy alone does not establish suitability for embedded deployment, where processing time, memory and power constraints must also be considered. Recent research has highlighted the importance of acoustic representation and classifier architecture in underwater acoustic target recognition~\cite{muller2024survey}. Previous work has demonstrated real-time acoustic processing using embedded hydrophone systems~\cite{toma2018smart}, while Vasconcelos et al.~\cite{vasconcelos2026maritime} investigated low-power maritime monitoring architectures that separate continuous vessel detection from classification. Park et al.~\cite{park2026embedded} further demonstrated lightweight ship-radiated sound classification on a Raspberry Pi, highlighting the importance of measured inference efficiency. These studies establish the potential of embedded acoustic monitoring and motivate the selection of classifiers that balance classification performance with hardware resource requirements. To select a suitable classifier, we must consider the computational cost of both acoustic preprocessing and model inference~\cite{vasconcelos2026maritime}. Comparing complete audio-to-prediction pipelines on the same device allows us to assess these costs alongside classification performance and identify suitable deployment trade-offs.

To investigate these trade-offs, we implemented and evaluated six underwater vessel classifiers on a Raspberry Pi 5, spanning linear models, nonlinear classifiers and neural networks. Using the ShipsEar benchmark~\cite{shipsear}, we compared classification performance, processing latency and hardware resource usage. Supporting experiments examined Short-Time Fourier Transform (STFT) configuration, statistical pooling and neural architecture choices to inform the design of classifiers for embedded deployment. The resulting comparison informed the selection of a radial basis function support vector machine (RBF-SVM) for deployment on a Raspberry Pi Zero 2 W. The main contributions of this work are:

\begin{itemize}

    \item Systematic embedded evaluation: We implement and benchmark six vessel classification models on a Raspberry Pi 5 under a unified audio-to-prediction pipeline, enabling direct comparison of classification accuracy, end-to-end latency, CPU utilisation, memory consumption, and monitored power.
    \item Representation and computational trade-off analysis: We investigate the impact of acoustic feature configurations and model architectures on classification performance and computational requirements. Our measurements identify feature extraction as the dominant processing cost for classical classifiers, identifying a practical target for reducing end-to-end latency.
\item Resource-constrained deployment validation: We demonstrate that an RBF-SVM achieves near-CNN classification accuracy with a median end-to-end processing latency approximately 76.7\% lower than that of the CNN on Raspberry Pi 5. Its deployment on Raspberry Pi Zero 2 W further demonstrates sufficient throughput for processing consecutive five-second audio segments in a resource-constrained environment.

\end{itemize}

\section{Methodology}

The ShipsEar dataset~\cite{shipsear} was selected because it has  five-second audio segments of underwater hydrophone recordings sampled at 16 kHz, covering five classes comprising four vessel groups and background noise. These recordings provide real-world representative acoustic inputs for investigating vessel classification in practical maritime monitoring scenarios. The experiments used 1,778 training and 445 test segments, with the same test set used across all six models.  Recording-level data leakage can inflate vessel classification performance~\cite{hashemi2026uniqueship}.  For this dataset,  source-recording identities could not be reconstructed for the frozen split, and recording-independent generalisation could not be verified. 

We adopted the Short-Time Fourier Transform (STFT) as the common acoustic representation for all classifiers. We evaluated different FFT sizes and frequency-bin configurations using a consistent one-layer Convolutional Neural Network (CNN) architecture. A baseline of 2048-point FFT with 64 frequency bins served as the reference, while our final configuration uses a 4096-point FFT with 128 bins. This established a common acoustic representation for benchmarking classification performance and computational requirements on the Raspberry Pi 5.

We selected the classifiers to compare linear and nonlinear models using pooled acoustic features with neural models operating directly on spectrograms. We used logistic regression and a linear Support Vector Machine (SVM) to establish computationally lightweight classification baselines. We then used a Radial Basis Function-SVM (RBF-SVM) and a Multilayer Perceptron (MLP) to investigate whether nonlinear modelling could improve classification performance sufficiently to warrant additional computational cost on embedded hardware. To reduce feature dimensionality, we applied temporal mean and standard deviation pooling, a statistical aggregation approach also used in acoustic representation learning~\cite{snyder2018xvectors}. The mean summarises the typical spectral response of each frequency bin, while the standard deviation captures its variation over time. Concatenating both statistics across the $128\times475$ spectrogram produces a 256-dimensional feature vector.

Building on the statistical classification approaches, we investigated CNNs operating directly on spectrograms, varying kernel geometry, pooling and dropout to develop
a compact two-layer architecture. Square kernels were examined for local time-frequency patterns, while rectangular kernels with greater temporal extent were
evaluated for longer-duration acoustic variation, following previous work on time-frequency convolutional kernel design~\cite{pons2017temporal}. We also applied temporal mean and standard deviation pooling to intermediate convolutional feature maps, aggregating each channel-frequency response across time before classification.

To complement convolution-based feature extraction, a compact patch transformer with self-attention is used to model relationships between acoustic features across different temporal intervals. Similarly to the Audio Spectrogram Transformer~\cite{ast_gong2021}, the spectrogram was divided into patches, projected into embeddings and processed using self-attention. Patch geometry is then investigated because the number and dimensions of patches influence both the acoustic information represented and the computational requirements of attention. Local $8\times20$ patches were compared with full-frequency $128\times20$ and $128\times10$ patches, producing 368, 23 and 47 patches, respectively. Full-frequency temporal patches~\cite{fftp_makineni2025} incorporate all 128 frequency bins within each temporal interval, allowing spectral components to be represented together while reducing the sequence length compared with local patching. Varying temporal widths were examined to investigate the balance between temporal resolution and the number of patches. 
To evaluate embedded performance, all six classifiers were implemented on a Raspberry Pi 5 using a shared audio-to-prediction pipeline. The CNN and Patch Transformer processed $1\times128\times475$ spectrograms, while the remaining four classifiers used 256-dimensional temporally pooled features. We trained all six classifiers across five random seeds and report their mean accuracy and macro-F1 with standard deviations. Logistic regression, linear SVM and RBF-SVM produced identical classification scores across these runs.


\begin{figure}[!h]
    \centering
    \includegraphics[width=0.8\columnwidth]
    {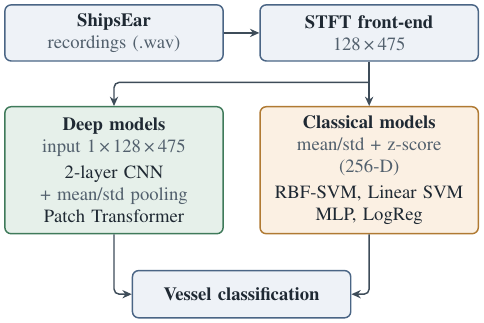}
    \caption{Raspberry Pi 5 underwater vessel classification pipeline. Five-second audio segments are converted into a $128\times475$ STFT representation. The CNN and Patch Transformer use spectrograms; the other four classifiers use 256-dimensional mean and standard deviation features.}
    \label{fig:pi_pipeline}
\end{figure}

 We measured the total time for the complete audio-to-prediction pipeline using the same 445 test segments.  Processing time was separated into audio loading, feature extraction and model inference, with median latencies reported for all six classifiers, with a 95th-percentile inference latency.  A separate 60-second sustained workload was used to examine throughput, CPU utilisation, memory consumption and processor temperature. Power usage was estimated using the Raspberry Pi 5's monitored Power Management Integrated Circuit (PMIC) rails. Process-level and system-wide measurements were recorded separately to characterise the hardware requirements of each classifier.

\section{Results and Discussion}

The Raspberry Pi 5 evaluation demonstrated that the relationship between classification performance and computational efficiency is not always straightforward, as more accurate or complex models do not necessarily incur higher inference latency. To understand these differences, we first examine the effects of feature configuration and architectural optimisation before comparing the classification performance, execution time and hardware requirements of the six deployed models.

We investigated the effect of STFT parameters on classification performance using a consistent one-layer CNN architecture. Increasing the FFT size from 2048 to 4096 and the frequency bins from 64 to 128 improved macro-F1 from 88.41 to 93.84 while increasing model parameters from 1,893 to 3,173. 

Table~\ref{tab:architecture_optimisation} presents the effects of statistical pooling on CNN classifier dimensionality and the influence of patch geometry on Transformer complexity and classification performance. These experiments investigate two approaches to improving neural classification while controlling model complexity, examining practical performance on embedded hardware.

\begin{table}[!htbp]
\centering
\caption{Effect of CNN statistical pooling and Transformer patch geometry on ShipsEar classification.}
\label{tab:architecture_optimisation}

\small
\setlength{\tabcolsep}{2pt}
\renewcommand{\arraystretch}{1.15}

\begin{tabular*}{\columnwidth}{@{\extracolsep{\fill}}lrrr@{}}
\hline
Configuration & Input / Patches & Params & Macro-F1 (\%) \\
\hline
\multicolumn{4}{l}{\textbf{CNN: Statistical pooling}} \\
Without pooling & 302,080 & $\sim$1,668,000 & 89.87 \\
Mean + std & 5,120 & 183,237 & \textbf{96.98} \\
\hline
\multicolumn{4}{l}{\textbf{Transformer: Patch geometry}} \\
Local ($8\times20$) & 368 & 201,349 & 74.53 \\
Full freq. ($128\times20$) & 23 & 464,389 & 96.85 \\
Full freq. ($128\times10$) & 47 & 303,621 & \textbf{97.11} \\
\hline
\end{tabular*}

\vspace{3pt}
\parbox{\columnwidth}{%
\small
\emph{Notes:} Input denotes CNN classifier features or
Transformer patch counts. CNN feature dimensions
are $80\times32\times118$ without pooling and
$160\times32$ with pooling. Results are from
exploratory experiments.
}
\end{table}

As shown in Table~\ref{tab:architecture_optimisation}, temporal statistical pooling reduced the CNN classifier input from 302,080 to 5,120 features and the overall
parameter count by approximately 89\%, while increasing macro-F1 from 89.87\% to 96.98\%. These exploratory results suggest that temporal aggregation retained
sufficient discriminative information while reducing the size of the fully connected classification head.

For the patch transformer, patch geometry influenced how spectral information was represented and processed.  The improvement in macro-F1 from 74.53\% to 96.85\% suggests that representing cross-frequency relationships within individual patches is beneficial for distinguishing vessel sounds. Although this configuration reduced the sequence from 368 to 23 patches, the parameter count increased, demonstrating that fewer tokens do not necessarily produce a smaller model. Furthermore, the exploratory $128\times10$ configuration achieved a slightly higher macro-F1 of 0.9711 with fewer parameters than the $128\times20$ configuration. 

In table~\ref{tab:pi_performance} we compare classification performance, model storage and execution time for all six models on the Raspberry Pi 5. Across five seeds, the two-layer CNN achieved a mean accuracy of $(99.06 \pm 0.33)\%$ and macro-F1 of $(99.14 \pm 0.32)\%$, compared with 98.88\% accuracy and 98.83\% macro-F1 for the RBF-SVM. Their median end-to-end latencies were 120.655~ms and 28.109~ms, respectively.

\begin{table}[!htpb]
\centering
\caption{ShipsEar classification performance and latency on Raspberry Pi 5.}
\label{tab:pi_performance}

\footnotesize
\setlength{\tabcolsep}{1pt}
\renewcommand{\arraystretch}{1.2}

\begin{tabular*}{\columnwidth}{@{\extracolsep{\fill}}lccrrr@{}}
\hline
Model &
\shortstack{Acc.\\(\%)} &
\shortstack{Macro-F1\\(\%)} &
\shortstack{Size\\(KiB)} &
\shortstack{Infer.\\(ms)} &
\shortstack{E2E\\(ms)} \\
\hline

CNN (2-layer)
& $\mathbf{99.06\pm0.33}$
& $\mathbf{99.14\pm0.32}$
& 723.08 & 91.050 & 120.655 \\

\shortstack[l]{Patch\\Transformer}
& $96.40\pm0.71$
& $96.48\pm0.63$
& 1824.11 & 2.776 & 29.738 \\

MLP
& $96.05\pm0.79$
& $96.08\pm0.71$
& 1179.03 & 0.901 & 27.772 \\

Linear SVM
& $90.34\pm0.00$
& $89.70\pm0.00$
& 17.25 & \textbf{0.614} & \textbf{27.179} \\

\shortstack[l]{Logistic\\regression}
& $87.87\pm0.00$
& $88.03\pm0.00$
& \textbf{12.39} & 0.652 & 27.270 \\

\textbf{RBF-SVM}$^\dagger$
& $98.88\pm0.00$
& $98.83\pm0.00$
& 2428.27 & 1.193 & 28.109 \\

\hline
\end{tabular*}

\vspace{3pt}
\parbox{\columnwidth}{%
\footnotesize
\emph{Notes:} Classification scores are mean $\pm$ sample SD
across five seeds on 445 test segments. Latencies are
medians; E2E includes audio loading, feature extraction
and inference. Bold indicates the best value per metric.
$^\dagger$ Selected for Raspberry Pi Zero 2 W deployment.
}
\end{table}

The RBF-SVM achieved a $4.29\times$ end-to-end speedup over the CNN, with only 0.31 percentage points lower macro-F1, supporting its selection for resource-constrained deployment. Examining the processing-time breakdown, median STFT feature-extraction times averaged 25.89 ms across the six classifiers, reflecting their shared preprocessing configuration. For the RBF-SVM, feature extraction accounted for approximately 93\% of end-to-end processing time, while inference required only 1.193 ms.  Its 95th-percentile (P95) inference latency was 1.243 ms, only 4.2\% above the median. This demonstrates that informative spectral features can support high classification accuracy with relatively inexpensive model inference.  In contrast, CNN inference accounted for approximately 75\% of its total processing time, making its slightly higher mean classification performance considerably more expensive on the Raspberry Pi 5. The Patch Transformer also exhibited much faster inference than the CNN despite having more parameters, although the cause of this difference remains unresolved without operator-level profiling. These results show that embedded deployment suitability depends on how the complete processing pipeline executes on the target hardware, over model complexity alone.

Beyond processing latency, Table~\ref{tab:pi_resources} shows the hardware resource requirements during sustained operation. Compared with the RBF-SVM, the CNN required 2.17 times greater system CPU utilisation, approximately 67\% more process memory and 34\% higher monitored PMIC power. The RBF-SVM maintained near-CNN classification accuracy with substantially lower hardware demands. The results provide a controlled comparison of classification performance and computational requirements under consistent benchmark conditions.


\begin{table}[!h]
\centering
\caption{Hardware resource utilisation on Raspberry Pi 5.}
\label{tab:pi_resources}

\small
\setlength{\tabcolsep}{3pt}
\renewcommand{\arraystretch}{1.12}

\begin{tabular*}{\columnwidth}{@{\extracolsep{\fill}}lrrrr@{}}
\hline
Model &
\shortstack{CPU\\(\%)} &
\shortstack{RAM\\(MiB)} &
\shortstack{PMIC\\(W)} &
\shortstack{Max. temp.\\($^\circ$C)} \\
\hline

CNN (2-layer)
& 54.56 & 490.52 & 5.193 & 54.50 \\

Patch Transformer
& 30.48 & 452.11 & 4.153 & 50.10 \\

MLP
& 25.16 & \textbf{287.25} & 3.960 & 50.15 \\

Linear SVM
& \textbf{25.15} & 291.19 & 3.895 & 52.35 \\

Logistic regression
& 25.56 & 290.69 & 4.014 & 52.35 \\

\textbf{RBF-SVM}
& 25.16 & 293.14 & \textbf{3.863}
& \textbf{48.50} \\

\hline
\end{tabular*}

\vspace{3pt}
\parbox{\columnwidth}{%
\small
\emph{Notes:} Measurements taken during 60-second
sustained execution. CPU is mean system utilisation;
RAM is process memory. PMIC power covers monitored
rails, not total input power. Bold indicates the
lowest value per metric.
}
\end{table}

\begin{figure}[!h]
    \centering
    \includegraphics[width=\columnwidth]{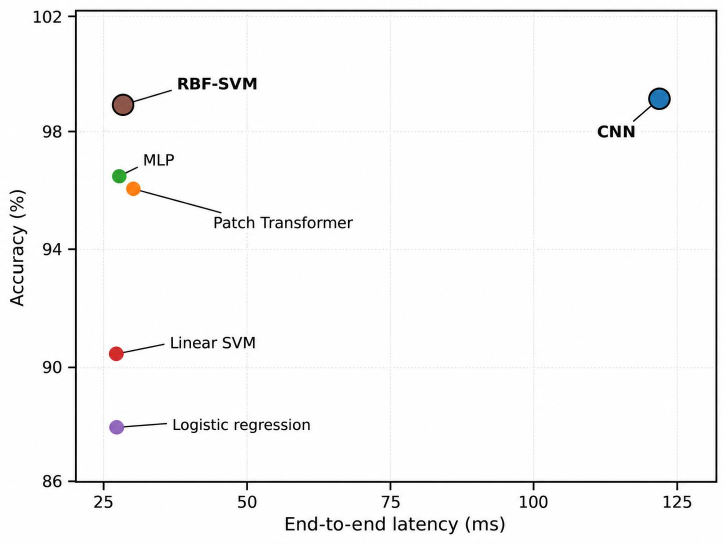}
    \caption{Classification accuracy versus median
    end-to-end latency for six classifiers on the
    Raspberry Pi 5. The RBF-SVM achieved 98.88\%
    accuracy compared with 99.06\% for the CNN,
    while providing a 4.29-fold speedup.}
    \label{fig:pi5_accuracy_latency}
\end{figure}

The processing time for the model on Raspberry Pi 5 further showed that model size does not directly determine computational cost.
As shown in Fig.~\ref{fig:pi5_accuracy_latency}, the RBF-SVM
achieved comparable accuracy to the CNN while requiring
4.29 times less end-to-end processing time.
This favourable accuracy-latency trade-off motivated its
selection for deployment on the more resource-constrained
Raspberry Pi Zero 2W.

\subsection{Raspberry Pi Zero 2 W Implementation}

\begin{table}[!htbp]
\centering
\caption{RBF-SVM processing performance and resource usage on Raspberry Pi 5 and Zero 2 W using ShipsEar.}
\label{tab:rpi_deployment}
\small
\renewcommand{\arraystretch}{1.15}

\begin{tabular*}{\columnwidth}{@{\extracolsep{\fill}}lrr@{}}
\hline
Metric & Pi 5 & Pi Zero 2 W \\
\hline
Feature extraction (ms) & 26.10 & 327.88 \\
SVM inference (ms) & 1.19 & 9.40 \\
Peak process RAM (MB) & 307.4 & 119.02 \\
RAM / device capacity (\%) & 7.7 & 24.4 \\
Mean process CPU (\%) & 25.07 & 25.25 \\
\hline
\end{tabular*}
\end{table}

Table~\ref{tab:rpi_deployment} compares feature extraction, model inference, and resource usage on both devices. On the Raspberry Pi Zero 2 W, the RBF-SVM maintained 98.88\% accuracy and 98.83\% macro-F1 across 445 ShipsEar test recordings. The device achieved a median end-to-end latency of 349.09~ms, including WAV loading, feature extraction, and inference, with a sustained throughput of 2.962 classifications/s. This exceeds the 0.2 classifications/s required for consecutive five-second recordings by 14.8 times. During inference across the full test set, we measured an average device power consumption of \textbf{1.17~W}, compared with \textbf{0.55~W} at idle, representing an additional \textbf{0.62~W} under load.

Previous ShipsEar studies reported accuracies of
99.87\% using ACNN--DRACNN (0.61M parameters)~\cite{ni2024acnn}, 99.19\% using CAPSE--ViT (1.90M)~\cite{najamuddin2025capse}, and 98.64\% using SSA--CACNN (0.26M)~\cite{ji2025ssacacnn}.
Our two-layer CNN achieved a mean accuracy of $(99.06 \pm 0.33)\%$ across five seeds using only 0.183M parameters on the adopted ShipsEar split. Our earlier investigation evaluated vessel classification on both ShipsEar and DeepShip, including recording-level partitioning on DeepShip~\cite{arxiv}. The present study extends this work through classification and hardware performance evaluation on Raspberry Pi devices.

\section{Conclusion}

Our findings demonstrate that careful acoustic feature selection and model optimisation can support accurate and computationally efficient underwater vessel classification on embedded hardware. Experiments with CNNs showed that optimising STFT parameters improved classification performance, while statistical pooling improved performance and reduced model complexity. Full-frequency Transformer patching also improved performance while reducing the number of patches, reinforcing the importance of architectural design.

The Raspberry Pi 5 evaluation demonstrated how these design choices translate into practical edge performance. Across five seeds, the CNN achieved the highest mean accuracy of $(99.06 \pm 0.33)\%$, while the RBF-SVM achieved a comparable 98.88\% accuracy with $4.29\times$ faster end-to-end processing and only 1.19~ms for classifier inference. These results demonstrate that classical classifiers, such as the RBF-SVM, supported by effective feature engineering, can approach the performance of deep learning models at substantially lower inference costs. This motivated its deployment on the more resource-constrained Raspberry Pi Zero 2 W, where it maintained the same classification accuracy and achieved 14.8 times the throughput required for consecutive five-second recordings. These results demonstrate the feasibility of CPU-only vessel classification on low-resource embedded hardware, supporting local acoustic monitoring while reducing reliance on continuous data transmission to shore.

\newpage
\bibliographystyle{IEEEtran}
\bibliography{references}

@article{shipsear,
  author = {Santos-Dom{\'{\i}}nguez, David and Torres-Guijarro, Soledad and Cardenal-L{\'{o}}pez, Antonio and Pena-Gimenez, Antonio},
  title = {{ShipsEar}: An underwater vessel noise database},
  journal = {Applied Acoustics},
  volume = {113},
  pages = {64--69},
  year = {2016},
  doi = {10.1016/j.apacoust.2016.06.008}
}

@misc{arxiv,
  author = {Soti, Abishek and Sone, Thura Pyae and Ibnul, Naqib and Aung, Htoo Htet and Zhong, Henry and Cohen, Gregory and Xu, Ying},
  title = {Towards Deployable Underwater Vessel Classification},
  year = {2026},
  eprint = {2609.29179},
  archivePrefix = {arXiv},
  primaryClass = {cs.SD},
  doi = {10.48550/arXiv.2609.29179},
  url = {https://arxiv.org/abs/2609.29179}
}

@inproceedings{ast_gong2021,
  author = {Gong, Yuan and Chung, Yu-An and Glass, James R.},
  title = {{AST}: Audio Spectrogram Transformer},
  booktitle = {Proceedings of Interspeech},
  pages = {571--575},
  year = {2021},
  doi = {10.21437/Interspeech.2021-698}
}

@article{underwater_networks_pal2022,
  author = {Pal, Amitangshu and Campagnaro, Filippo and Ashraf, Khadija and Rahman, Md Rashed and Ashok, Ashwin and Guo, Hongzhi},
  title = {Communication for Underwater Sensor Networks: A Comprehensive Summary},
  journal = {ACM Transactions on Sensor Networks},
  volume = {19},
  number = {1},
  articleno = {22},
  pages = {1--44},
  year = {2022},
  doi = {10.1145/3546827}
}

@misc{fftp_makineni2025,
    title={Full-Frequency Temporal Patching and Structured Masking for Enhanced Audio Classification},
    author={Makineni, Aditya and Geng, Baocheng and Tian, Qing},
    year={2025},
    eprint={2508.21243},
    archivePrefix={arXiv},
    primaryClass={cs.SD},
    doi={10.48550/arXiv.2508.21243}
}

@incollection{pierpoint2026soundscapes,
  author = {Pierpoint, Chris and Morrish, James and Hedgeland, David},
  title = {Real-Time Remote Monitoring of Underwater Soundscapes},
  booktitle = {The Effects of Noise on Aquatic Life IV},
  publisher = {Springer},
  year = {2026},
  doi = {10.1007/978-3-031-94229-7_183-1}
}

@article{park2026embedded,
  author = {Park, Sangwon and Kim, Dongjun and Byun, Sung-Hoon and Park, Sangwook},
  title = {Ultra-Lightweight Ship-Radiated Sound Classification for Real-time Embedded Inference},
  journal = {IEEE Embedded Systems Letters},
  year = {2026},
  doi = {10.1109/LES.2026.3708295}
}

@article{ni2024acnn,
    author = {Ni, Junshuai and Ji, Fang and Lu, Shaoqing and Feng, Weijia},
    title = {An Auditory Convolutional Neural Network for Underwater Acoustic Target Timbre Feature Extraction and Recognition},
    journal = {Remote Sensing},
    volume = {16},
    number = {16},
    pages = {3074},
    year = {2024},
    doi = {10.3390/rs16163074}
}

@article{najamuddin2025capse,
    author = {Najamuddin and Sheikh, Usman Ullah and Sha'ameri, Ahmad Zuri},
    title = {{CAPSE-ViT}: A Lightweight Framework for Underwater Acoustic Vessel Classification Using Coherent Spectral Estimation and Modified Vision Transformer},
    journal = {Archives of Acoustics},
    volume = {50},
    number = {2},
    pages = {161--171},
    year = {2025},
    doi = {10.24425/aoa.2025.153662}
}

@article{ji2025ssacacnn,
    author = {Ji, Fang and Lu, Shaoqing and Ni, Junshuai and Li, Ziming and Feng, Weijia},
    title = {Underwater Target Recognition Method Based on Singular Spectrum Analysis and Channel Attention Convolutional Neural Network},
    journal = {Sensors},
    volume = {25},
    number = {8},
    pages = {2573},
    year = {2025},
    doi = {10.3390/s25082573}
}

@article{toma2018smart,
    author = {Toma, Daniel Mihai and Masmitja, Ivan
              and del Rio, Joaquin and Martinez, Enoc
              and Artero-Delgado, Carla and Casale, Alessandra
              and Figoli, Alberto and Pinzani, Diego
              and Cervantes, Pablo and Ruiz, Pablo
              and Meme, Simone and Delory, Eric},
    title = {Smart Embedded Passive Acoustic Devices
             for Real-Time Hydroacoustic Surveys},
    journal = {Measurement},
    volume = {125},
    pages = {592--605},
    year = {2018},
    doi = {10.1016/j.measurement.2018.05.030}
}

@article{vasconcelos2026maritime,
    author = {Vasconcelos, Hugo Mesquita
              and Sousa, Pedro J. S. C. P.
              and Dias, Susana and Pinto, Jose P.
              and van Golde, Ilmer D.
              and Tavares, Paulo J.
              and Moreira, Pedro M. G. P.},
    title = {A Low-Power Architecture for Passive
             Acoustic Autonomous Maritime Surveillance},
    journal = {Journal of Marine Science and Engineering},
    volume = {14},
    number = {9},
    pages = {815},
    year = {2026},
    doi = {10.3390/jmse14090815}
}

@article{muller2024survey,
    author = {M{\"u}ller, Nils and Reermann, Jens
              and Meisen, Tobias},
    title = {Navigating the Depths: A Comprehensive Survey
             of Deep Learning for Passive Underwater
             Acoustic Target Recognition},
    journal = {IEEE Access},
    volume = {12},
    pages = {154092--154118},
    year = {2024},
    doi = {10.1109/ACCESS.2024.3480788}
}

@misc{hashemi2026uniqueship,
    author = {Hashemi, Connor and Stout, Trevor
              and Hoogs, Anthony and Parham, Jason},
    title = {{UniqueShip}: Mitigating Data Leakage in
             Acoustic Ship Classification Benchmark Datasets},
    year = {2026},
    eprint = {2609.13659},
    archivePrefix = {arXiv},
    url = {https://arxiv.org/abs/2609.13659}
}

@inproceedings{pons2017temporal,
  author    = {Pons, Jordi and Serra, Xavier},
  title     = {Designing Efficient Architectures for Modeling
               Temporal Features with Convolutional Neural Networks},
  booktitle = {2017 IEEE International Conference on Acoustics,
               Speech and Signal Processing (ICASSP)},
  pages     = {2472--2476},
  year      = {2017},
  doi       = {10.1109/ICASSP.2017.7952601}
}

@inproceedings{snyder2018xvectors,
  author = {Snyder, David and Garcia-Romero, Daniel and Sell, Gregory and Povey, Daniel and Khudanpur, Sanjeev},
  title = {{X-Vectors}: Robust {DNN} Embeddings for Speaker Recognition},
  booktitle = {2018 IEEE International Conference on Acoustics, Speech and Signal Processing (ICASSP)},
  pages = {5329--5333},
  year = {2018},
  publisher = {IEEE},
  doi = {10.1109/ICASSP.2018.8461375}
}

\end{document}